\documentclass[sigconf]{acmart}
\renewcommand\footnotetextcopyrightpermission[1]{}
\usepackage{booktabs} 
\usepackage{balance}
\usepackage{enumitem}
\usepackage{bm,array}
\usepackage{dblfloatfix}
\usepackage{verbatim}
\usepackage{url}
\usepackage{tabularx}
\usepackage{pgfplots}
\usepackage{caption}
\usepackage{subcaption}
\pgfplotsset{compat=1.13}
\setcopyright{none}
\acmConference[VAMS workshop @ RecSys '17]{ACM RecSys}{August 2017}{Como, Italy}

\renewcommand\footnotetextcopyrightpermission[1]{}

\newcolumntype{C}{>{\centering\arraybackslash}p{3em}}
\begin{document}
\title{Price and Profit Awareness in Recommender Systems}
\author{Dietmar Jannach}
\affiliation{%
	\institution{TU Dortmund, Germany}
}
\email{dietmar.jannach@tu-dortmund.de}
\author{Gediminas Adomavicius}
\affiliation{%
  \institution{University of Minnesota, USA}
}
\email{gedas@umn.edu}

\begin{abstract}
Academic research in the field of recommender systems mainly focuses on the problem of maximizing the users' utility by trying to identify the most relevant items for each user. However, such items are not necessarily the ones that maximize the utility of the service provider (e.g., an online retailer) in terms of the business value, such as
profit. One approach to increasing the providers' utility is to incorporate purchase-oriented information, e.g., the price, sales probabilities, and the resulting profit, into the recommendation algorithms. In this paper we specifically focus on price- and profit-aware recommender systems. We provide a brief overview of the relevant literature and use numerical simulations to illustrate the potential business benefit of such approaches.
\end{abstract}

\maketitle
\section{Introduction}
Academic research on recommender systems (RS) has largely focused on the consumer's viewpoint, i.e., often the main goal 
is to design systems that help users find relevant items, services, or information.  From the business perspective, providing a functionality that is oriented towards helping consumers (e.g., site visitors) can be valuable in itself, e.g., lead to higher customer loyalty and retention. 
However, businesses are increasingly looking to recommenders as tools that can play a more direct role in improving business-related measures, such as sales, profit, and customer involvement \cite{JannachResnickEtAl2016}.  Such profit-oriented goals  may revolve around the capability of an RS to \emph{influence} the behavior of the users, e.g., by stimulating more or different sales or by keeping the users involved with the service. These profit-oriented goals can, however, easily be in conflict with a consumer's needs, since the recommendation service might no longer be simply suggesting items with the expected highest \emph{utility} for the consumer 
but rather explicitly taking into account the firm's own business-oriented considerations.  Therefore, achieving the proper balance between the two sides constitutes an important and interesting research question. {\let\thefootnote\relax\footnote{Presented at the 2017 Workshop on Value-Aware and Multi-Stakeholder Recommendation (VAMS) collocated with ACM RecSys 2017.}}

There is a wide variety of potential ways that a recommender can take a provider's viewpoint into account.  One of the more natural, direct approaches is to incorporate some direct purchase-oriented information -- such as the price of a recommended item, the probability of a purchase of an item (at the given price), and the profit resulting from a purchase -- into the recommendation algorithm. In this paper we therefore focus on the specific topic of \emph{price and profit awareness in recommender systems} and provide a brief, consolidated overview and discussion of relevant literature, ideas, challenges, and future directions.

\label{sec:business-impact}

\subsection{Purposes and Revenue Models of RS}
There is a number of different ways a recommender system can create value for the provider, e.g., by influencing the users' navigation and purchase behavior, by increasing the ``discoverability'' of long-tail items, by increasing user engagement and loyalty, or simply by making the site easier or more entertaining to use \cite{JannachAdomavicius2016purpose}.
When designing an RS that takes the business-oriented perspective into account it is however not only important to know the intended purpose of the system, but also to understand the business and revenue models of the service.

In early days of recommender systems research, Resnick and Varian sketched some business and revenue models for \emph{stand-alone} recommendation services \cite{Resnick:1997:RS:245108.245121}.
Today, twenty years later, stand-alone recommendation services (e.g., MovieLens) are comparably rare.  More often, recommender systems are integrated components of an online service, e.g., of a social network or a online retail site.  The revenue model of the recommendation service is, therefore, often intricately linked with the overall revenue model of the site.

In \cite{PYChenHicss09Community}, Chen et al.~analyzed different revenue models of recommender systems.
According to their study, the main revenue indicators (i.e., effectiveness measures) that can be attributed to recommenders are: more sales, fee-based revenue through more transactions and subscriptions, and increased income through different types of other fees for, e.g., advertisements, sponsorship, or user referral, as well as for software and content licenses.
Most of the approaches to price- and profit-aware recommendation discussed in the next section take a short-term view and consider ``improved sales'' (including shifting sales to more profitable items) as the main revenue-based measure that a recommender should support.  Long-term approaches, while not the focus of the work presented in \cite{PYChenHicss09Community}, can however also be considered when designing recommendation services, e.g., by using measures like ``customer lifetime value'' (see Section \ref{sec:long-term}) or by incentivizing positive long-term provider behavior (and, thus, positive consumer experience) by down-ranking offers from non-competitive, restrictive, and customer-unfriendly providers for a significant period of time \cite{Krasnodebski:2016:CSR:2959100.2959124}.

\subsection{Assessing the Business Value of RS}
Assessing the business value of an RS is a challenging task, especially in academic settings, i.e., without direct access to a sales or production environment.
A number of academic studies explore the different ways of how recommenders impact users, e.g., in terms of their decision making and choice processes \cite{Senecal2004159,Haubl:2000:CDM:767773.768989}. Many of them are based on user studies and often focus on the users' perception of recommenders, e.g., investigating if the participants found it easier to make choices or if they were more confident in their decisions afterwards. To assess the business value at least to some extent, some studies analyze self-reported behavioral intentions of the participants and ask for their ``intention to return'' or ``intention to purchase'', e.g., after interacting with the system \cite{NilashiJannachEtAl2016}. Such studies can be helpful indicators about the value of a recommender, even though in many cases these user studies do not involve any actual purchase decisions.

A different approach to assess the impact of recommenders on customers is to analyze available real-world data and reconstruct the possible effects based on aggregated sales figures or user behavior data. For example, \cite{Oestreicher-Singer:2012:VHD:2398302.2398303} analyzed how often certain items are recommended (together with others) on Amazon.com, used the sales rank information of this platform to estimate demand levels, and thereby indirectly assessed the effects of the recommendations on product sales.  A similar
analysis was done in an earlier study in \cite{ChenWuYoon:04a}. A very different ex-post analysis approach was taken in \cite{DBLP:conf/icis/AdamopoulosT15}, where the authors use an established econometric method for demand modeling with aggregated data in order to investigate which factors contribute to the success of a recommendation and how strong these effects are. 

Differently from these approaches, one can also try to analyze the underlying recommendation-related phenomena theoretically, using analytical (mathematical) modeling and simulation.  Even though various simplifying assumptions are typically needed to make such analyses tractable, the resulting models and simulations can still provide rich, informative insights.  For example, \cite{Fleder:2009:BCN:1538966.1538968} use a stylized analytical model and comprehensive simulations to demonstrate how recommender systems can reduce sales diversity.

Finally, a limited number of research works report the outcomes of field studies (A/B tests), where the business-related measures include the change in actual sales, conversion rates, or click-through rates \cite{Dias:2008:VPR:1454008.1454054,DBLP:conf/recsys/JannachH09,Garcin:2014:OOE:2645710.2645745,DBLP:conf/pkdd/KirshenbaumFD12}.  Some of this work emphasizes the fact that the standard predictive accuracy measures often used in academic research are not necessarily good predictors of the online, real-world success of a recommendation algorithm.

\subsection{Potentially Negative Impacts of RS}
While many research studies focus on demonstrating the positive effects of recommenders, there are some explorations of the potential (and possibly unintended) harms associated with recommender systems. In particular, in cases when a recommender is embedded in a site, consumers could possibly get the impression that the recommendations are merely there to promote certain items. Due to the attribution theory, the perceived fairness of the recommendations might be limited \cite{PYChenHicss09Community} and can lead to lower trust toward the provider as a whole.

One of the few studies in that direction is described in \cite{Chau:2013:EEM:2747904.2748214}, where the authors found
that irrelevant or biased recommendations had significant negative effects on the trust in the recommender system or in the provider.  In addition, the work in \cite{Adomavicius2012} demonstrates a robust and strong side-effect of recommender systems on consumers' economic behavior in that inaccurate recommendations can distort consumers' willingness to pay. Finally, the analysis in \cite{Zhang:2013:PRM:2507157.2507167} based on portfolio theory indicates that it is sometimes better not to personalize the recommendations for certain users and to just recommend popular items.

Generally, considering that one use of recommenders is to stimulate cross-selling, the marketing literature suggests that enticing cross-buying behavior from certain types of customers (e.g., those that exhibit promotion purchase behavior) can even be unprofitable for businesses \cite{doi:10.1509/jm.10.0445}.
Another study \cite{fitzsimons2004reactance} revealed reactant (contrary) behavior of the recommender system users in cases when recommendations are in conflict with initial user impressions.

There are also other potentially undesirable effects of recommenders on customer purchase behavior. Recommenders can, for example, lead to ``rich-get-richer'' (blockbuster) effects, where sales concentrate on a few very popular items, leaving the potential of increased sales of long-tail items untapped \cite{Fleder:2009:BCN:1538966.1538968,JannachLercheEtAl2015}.
Furthermore, on marketplaces with competing providers, recommenders can have both positive and negative effects. While they can increase the market by making items known to consumers, they can also lead to a price competition and lower profits for providers \cite{BoonOrBane2016}.

\section{Price- and Profit-Aware Recommendation Approaches}
\label{sec:categorization}
Sales prices and profit margins are key elements of the firm's success, and in many business environments they have a direct effect on the demand (market size).  Given the evidence that recommenders are effective in influencing the user behavior in many domains, price and profitability should be considered when designing a recommender. Research on this topic is, however, relatively scarce and scattered.  Therefore, in this section we provide a brief overview of existing studies on price- and profit-aware recommender systems.

\subsection{Price-Sensitivity in the User Model}
Different consumers can have different price preferences, and even the preferences of an individual consumer can vary across product categories. To maximize the probability that a user accepts a recommendation (leading to a purchase), it can therefore be advantageous to consider the individual price preferences in the user model of the RS. For example, \citet{Ge:2014:CCF:2576772.2559169} propose methods to incorporate observable and unobservable cost factors into different latent factor models. An evaluation in the travel and tourism domain indicates that explicitly considering cost factors is advantageous in terms of recommendation accuracy when compared to plain latent factor models, which capture the user's price preferences only implicitly.

The work in \cite{Ge:2014:CCF:2576772.2559169} focused on long-term, category-independent price preferences of users. A recent customer purchase behavior analysis for a fashion retailer furthermore showed that considering price preferences in the context of a consumer's short-term shopping goal can be advantageous \cite{JannachSAC2017ECommerce}. Specifically, recommending items that have about the same price level of other items viewed in the current session more than doubled the recommendation-to-purchase conversion rate.

Finally, \cite{Chen:2014:PRM:2600428.2609608} investigated the use of consumer price preferences for \emph{transfer learning}. The authors tested different ways of incorporating price information into the recommendation process and showed that considering the price preferences is particularly helpful when customers make purchases in product categories that they have not explored before.

\subsection{Balancing Relevance and Profitability}

\subsubsection{Motivation} To illustrate the potential trade-off between recommendation relevance and profit optimization and to further emphasize the value of analytical modeling and simulation, we conducted the following experiment. We took the MovieLens 1M data set and assigned profitability values for each of the movies. The values were chosen randomly using a Gaussian distribution with a mean value of \$2 and defined minimum and maximum profit (\$0 and \$4, respectively).
In the context of the top-10 recommendation task, we then made different experiments with the recommendation re-ranking method proposed in \cite{DBLP:journals/tkde/AdomaviciusK12}, which was originally designed to balance accuracy and aggregate diversity but was modified by us to balance accuracy and profitability. The method takes the recommendations of an underlying baseline technique, in our case a matrix factorization algorithm,
and then greedily re-ranks items for each user according to their profitability. To avoid strong accuracy deteriorations, a threshold $T_R$ can be set with respect to the minimum allowable predicted rating, i.e., only items that surpass a certain rating value can appear in the top-10 list.\footnote{Other general-purpose re-ranking techniques such as \cite{JugovacJannachLerche2017eswa} can be applied as well.}

\begin{figure}[h!t]
    \centering
    \includegraphics[trim=80pt 125pt 70pt 115pt, clip, width=0.47\textwidth]{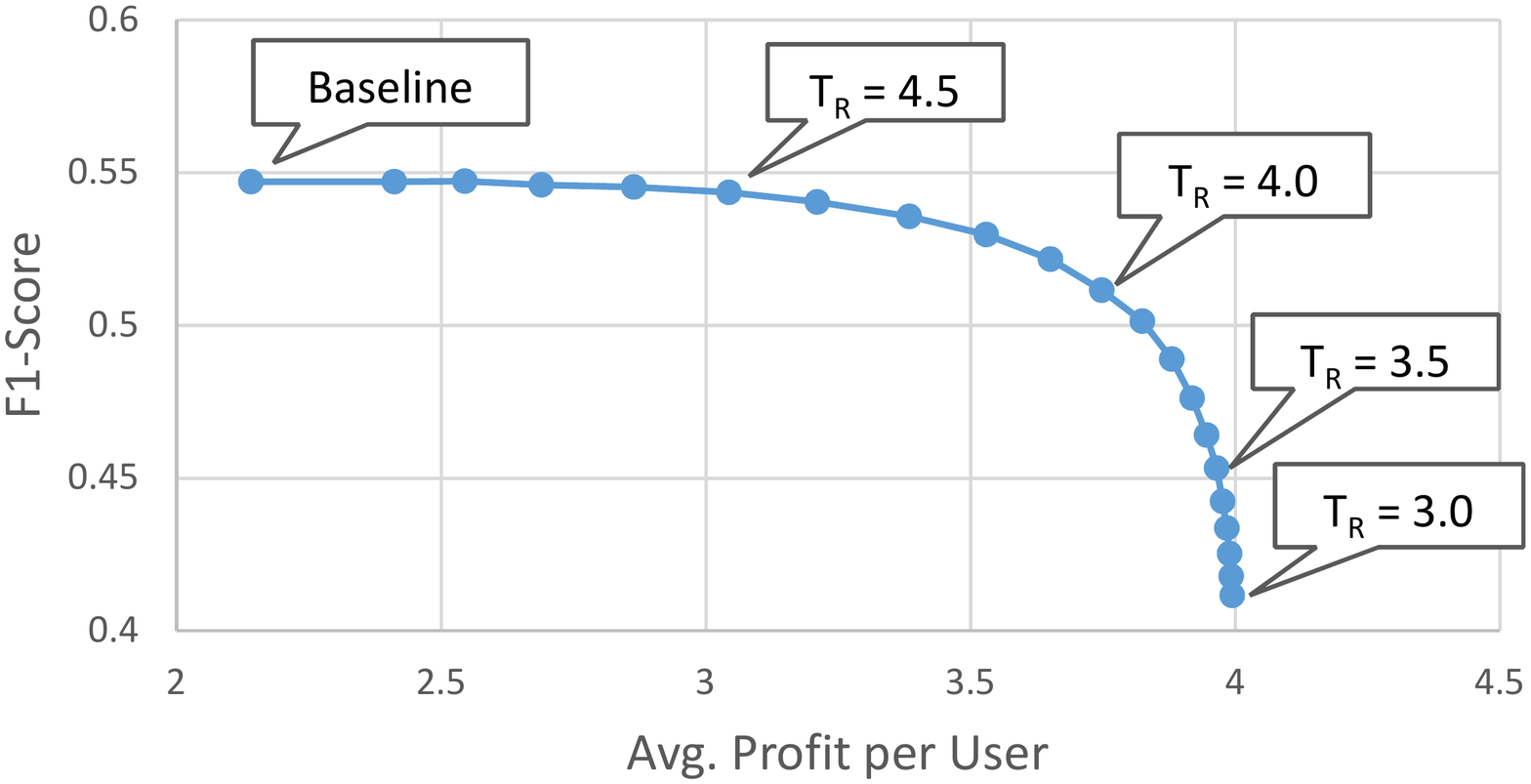}
	\vspace{-10pt}
    \caption{Profit-relevance tradeoff (guaranteed purchase)}
	\vspace{-5pt}
    \label{fig:numerical-experiment}
\end{figure}

The curve in Figure~\ref{fig:numerical-experiment} shows the obtained accuracy and profitability outcomes for different threshold values assuming that each user will buy one item from their top-10 list. The results show that a substantial profitability increase of about 50\% (i.e., from about 2\$ to 3\$) can be obtained with only a minimal loss in accuracy by lowering the predicted rating threshold to around 4.5.  When we further lower the threshold, e.g., allow items to be re-ranked that have a predicted score of 4, the accuracy loss is still lower than 10\%, but the increase in average profitability is at over 80\%.

\begin{figure}[h!t]
    \centering
    \includegraphics[trim=70pt 295pt 70pt 330pt, clip, width=0.47\textwidth]{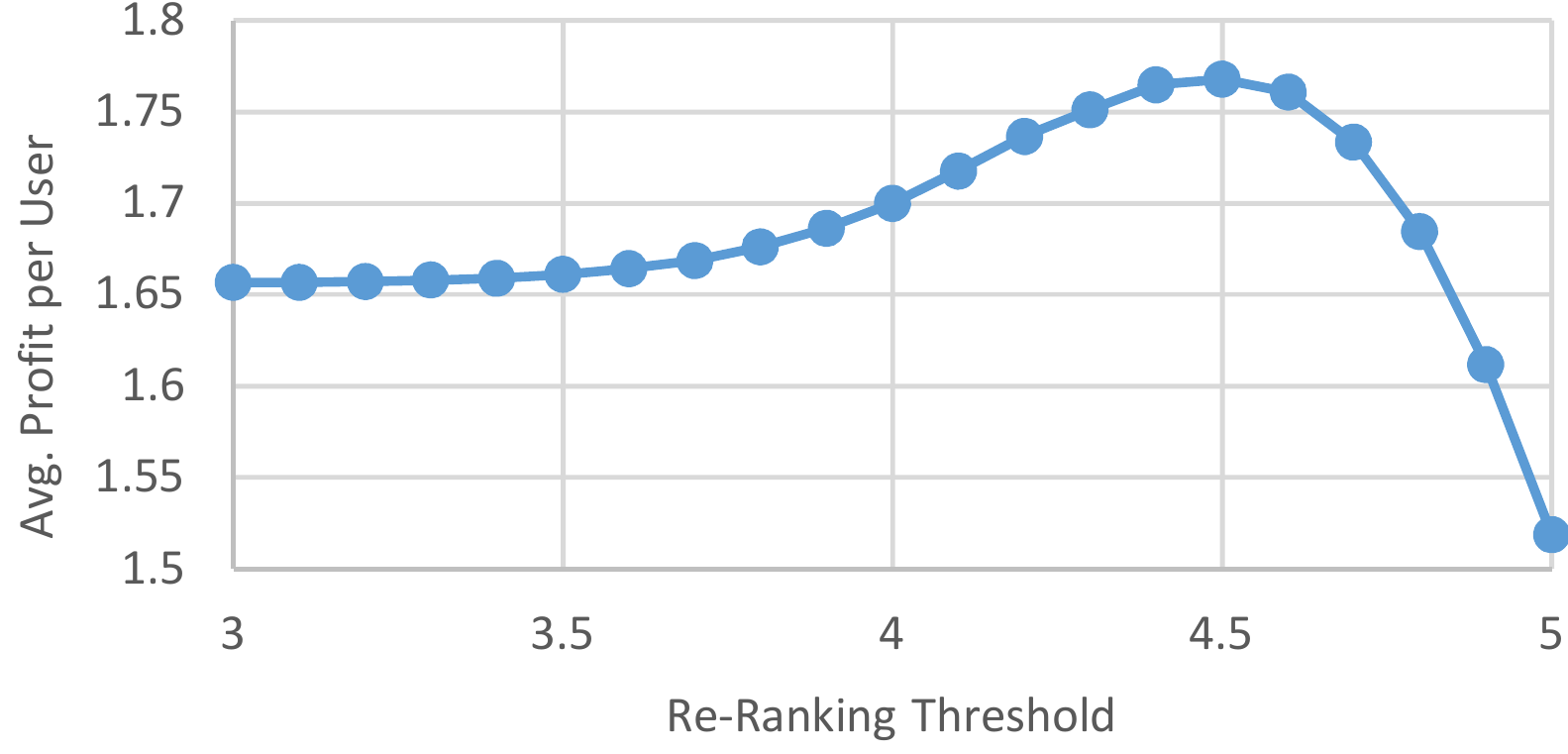}
    	\vspace{-10pt}
\caption{Profit optimization (relevance-based purchase)}
    	\vspace{-5pt}
    \label{fig:profit-modeling}
\end{figure}

Although Figure~\ref{fig:numerical-experiment} nicely illustrates the profit-relevance trade-off, its profitability analysis is based on the simplifying assumption that the users will always buy one item from their top-10 list.  However, if the recommender system produces not the most relevant (but more profitable) recommendations, in the real-world typically there would be some consequences for that.  E.g., there is some probability that a user may 
choose not to buy any of the items in the top-10 list, thus, resulting in a profit of 0 from that user.  Using a simple model of relevance-based purchasing, where the probability of purchase exponentially decays as the predicted item rating (i.e., item relevance) gets lower, Figure~\ref{fig:profit-modeling} shows that there is an optimal threshold from the profit-awareness perspective.  In particular, as the re-ranking threshold gets lowered to 4.5, more items become available for recommendation, among which more profitable items could be selected into top-10 lists.  Furthermore, the items are still highly relevant (predicted rating above 4.5), thus, the purchase likelihood is still fairly high.  However, as the re-ranking threshold gets below 4.5, the ability to choose from increasingly more items cannot overcome the rapidly decaying purchase probability, as the items become less relevant.  We also note that, without any profit-aware re-ranking (i.e., recommending the top-10 highest predicted items), the resulting average profit would be 1.30, i.e., lower than that of \emph{any} profit-aware setting shown in Figure~\ref{fig:profit-modeling}.

Overall, while the aforementioned results are based on a simulation with synthetic profit data and highly stylized, simplified assumptions about consumer purchase behavior, they can be seen as strong indicators of the economic potential of including profitability considerations when recommending.

\subsubsection{Incorporating Profit Information} As mentioned earlier, the items that are most relevant for the user often may not be the same as the ones with the best \emph{business value} for providers.  Therefore, several studies proposed approaches to balancing these potentially conflicting goals, using a wide variety of methodologies.

One early approach relies on a graph-based structure that captures the relationships between items, users, and their properties and use a random walk model to make recommendations \cite{conf/sdm/Brand05}. One specific property of the model is that it can be configured to nudge users to the more profitable states early during the random walk.

Several studies directly use optimization heuristics to balance accuracy and profitability.  In particular, \cite{Chen20081032} proposes recommendation strategies that combine purchase probability and product profitability factors to obtain an average (expected) margin, based on which the recommended items are ranked. Experiments on a synthetic dataset of shopping transactions show that an increase in profitability can be achieved without significant sacrifices in recommendation accuracy.

While the work in \cite{Chen20081032} is based on a heuristic scoring model, \cite{Wang20097299} frames the problem of balancing margin and item relevance as a mathematical optimization problem in which also side constraints (e.g., consumer budget) and other decision factors (e.g., customer satisfaction levels) can be integrated. Different optimization goals can be specified, to either optimize profitability or to generate a win-win situation for providers and consumers.
The use of mathematical optimization was also proposed in \cite{DBLP:journals/corr/abs-0908-3633,ValuePick2010,Hammar:2013:UMC:2507157.2507169,Azaria:2013:MRS:2507157.2507162}. While the work in \cite{DBLP:journals/corr/abs-0908-3633} is based on a number of simplifying assumptions and does not provide an empirical evaluation of the approach, the authors of \cite{Azaria:2013:MRS:2507157.2507162} validated the (short-term) effectiveness of their profit-maximization method through a user study on Mechanical Turk. Also, an optimization-based approach for the link recommendation problem on social networks has been proposed in \cite{ValuePick2010}, where ``profit'' for the provider was operationalized via a non-monetary metric. \citet{Hammar:2013:UMC:2507157.2507169} consider a non-personalized recommendation scenario and show that it is more profitable to recommend items with high purchase probability than to recommend best-sellers and proposes a corresponding greedy maximization procedure. The optimization-based approach in \cite{Lu:2014:SMM:2733085.2733086} takes into account yet another set of additional factors such as saturation effects, capacities, or competition amongst products. The authors furthermore optimize the revenue over a finite time horizon, aiming to find the optimal point in time to recommend certain items.

To further investigate the relationship between short-term profit maximization, relevance ranking, and trust, \citet{Panniello201687} conducted a field study in which they tested different
strategies. The results of their email-based study show that balancing relevance and profitability leads to higher profits than a simple content-based recommendation approach.  While this is somewhat expected, it also turned out that the profit-maximizing recommendations did not immediately lead to lower consumer trust.

A very different way of thinking about profit maximization for recommenders was put forward in the marketing field in \cite{doi:10.1509/jmkr.45.1.77}. Specifically, the question for a marketer is to assess if a product should be recommended -- given a limited number of recommendations that can be made -- even if it is likely to be purchased by a customer without the recommendation. The author argues that the decision for a marketing intervention (recommendation) should be made based not only on purchase probabilities, but also on the customers' expected behavior when they receive no recommendation.

\subsubsection{Combining Sales Promotion, Pricing, and Recommendations}
Aside from the ability to take into account profit information as additional input to the traditional item recommendation process, recommenders can also serve as a key instrument for businesses to \emph{proactively} promote certain items at certain prices, e.g., the promotional effect of recommendations could be further increased by offering items at a discounted price.  An important issue is whether consumers perceive recommendations more like advertisements than personalized buying suggestions, when the recommended items are explicitly labeled as being on sale. In particular, customers might be less attracted to the recommendations due to the perception of their ``promotional'' purpose. A recent analysis in \cite{JannachSAC2017ECommerce} showed the opposite for the case of an online fashion retailer. When an item on a recommendation list was presented as being discounted, that significantly increased the probability of the recommendation-to-purchase conversion for the item.

While in the analysis in \cite{JannachSAC2017ECommerce} the discounts were the same for all customers, some studies consider dynamic (customer-specific) prices to promote items in the context of recommendations. For example, \cite{Backhaus2010} proposes a collaborative filtering recommendation method that estimates the customer's willingness to pay (WTP) and groups customers into different segments. During the recommendation process, items can then be filtered out when the expected WTP of the customer is not positive.

The computation of personalized prices based on the estimated WTP was also the focus in \cite{Zhao:2015:ERP:2792838.2800178}. In their analysis, a number of factors were identified as influential on the customers' WTP for an item, e.g., the item's brand or the item's average rating. An experimental evaluation showed that the ability to successfully predict the WTP and to adapt the prices can lead to significant increases in profitability. However, the impact of such dynamic pricing on \emph{long-term} customer satisfaction has not been studied.

An integrated model that considers both the optimization of price discounts and the items to recommend has also been considered \cite{Jiang2015257,journals/ijitdm/JiangL12}. The general assumption in these studies is that a retailer may want to attract customers through a discount for a promoted item and then use recommendations to increase sales of non-discounted items. Numerical experiments indicate that the optimized online promotions can lead to a significantly increased overall profitability, even if no profits are made with the promoted product.

The work in \cite{Garfinkel200861} investigates the advantages of combining recommendations, sales promotions, and product bundles from the perspective of the price sensitive customer. They propose a ``shop bot'' (comparison shopping agent) system that determines the optimal deal for the customer using integer programming models.

Finally, an approach considering both the retailer's profit and the customer's savings was proposed in \cite{Jiang:2011:OEP:2020296.2020309}. Bundle recommendations are a common promotion mechanism on e-commerce sites, and in their work the authors focus on dynamically determining prices for such \emph{bundles}. Numerical analyses of their method show that optimizing the prices can lead to win-win situations for both sides.

\subsection{Towards the Long-Term Perspective}
\label{sec:long-term}
The approaches discussed so far focused on the \emph{short-term} perspective, i.e., incorporating price and profit awareness for the purposes of calculating the next recommendation.  The long-term perspective, albeit no less important, has been largely underexplored.

Some temporal dynamics are considered in \cite{Wang:2011:UMN:2009916.2010050}, where the modeling proposed by the authors is motivated by theoretical insights from consumer behavior theory. The assumption is that the marginal utility for consumers can vary over time, and the authors show how this aspect can be considered, e.g., by a matrix factorization recommendation algorithm.

More generally, the success of many businesses depends on the long-term loyalty and trust of the customers. Taking an analytical modeling approach to understand longer-term effects, Hosanagar et al.~\cite{HosanagarICIS2008} investigated the 
design of recommenders and confirmed that the best recommendation policies have to balance margin and relevance. Their results also indicated that the best policy can depend on the reputation status of the recommender and that in a low-reputation situation it is better to sacrifice profitability to restore reputation as a first goal.

Finally, the expected \emph{customer lifetime value} (CLV) has been investigated for many years in the management and marketing literature, see, e.g. \cite{Bhaduri2016}.
However, only few studies consider the expected CLV and related marketing activities like cross-buying, promotions, and reacquisition or retention pricing \cite{Blattberg2009157} in the context of recommender systems.  In one-to-one marketing, one of the common ways to assess the CLV is to use recency, frequency, and monetary (RFM) characteristics of customers. In particular, \cite{Liu2005181,Shih2008350} propose to group customers that have similar CLV values (according to an RFM model) into segments. Different weighted strategies can then be applied as part of recommendation process, e.g., by learning association rules per customer segment. However, the focus of the experimental evaluations in these studies was on current recommendation quality and not on the long-term value of the recommendations.

\section{Conclusions}
\label{sec:outlook}
From a business perspective, recommender systems provide new opportunities and means to implement known-to-be-effective marketing strategies like individual-level targeting or dynamic pricing \cite{doi:10.1509/jmkr.46.2.207,doi:10.1287/mnsc.1050.0373} and, more generally, to incorporate various price- and profit-related information into the recommendation process and to balance customers' needs with providers' goals.  However, in terms of recommender systems development, there still exists a significant gap between research done in technology-oriented fields like computer science and information systems and business-oriented fields like management science, consumer behavior, and marketing  \cite{Michalis2013}.  This paper attempts to bridge this gap by providing an overview of various studies on related topics that are scattered across different research areas. Price and profit awareness constitutes an important and promising research direction for recommender systems, and we hope that this overview will help stimulate further research.


\balance
 \bibliographystyle{abbrvnat}
\bibliography{literature-short}
\end{document}